**Title:** Vision Transformers increase efficiency of 3D cardiac CT multi-label segmentation

Authors: Lee Jollans*[a,b], Mariana Bustamante[a,b,d], Lilian Henriksson[b,c], Anders Persson[b,c], Tino Ebbers[a,b]

a: Department of Health, Medical and Caring Sciences, Linköping University, SE-581 83 Linköping, Sweden
b: Center for Medical Imaging and Visualization, Linköping University, SE-581 83 Linköping, Sweden
c: Department of Radiology, and Department of Health, Medicine and Caring Sciences, Linköping University, SE-581 83 Linköping, Sweden
d: deCODE Genetics/Amgen Inc., Sturlugata 8, 101 Reykjavik, Iceland

* corresponding author

E-mail addresses: lee.jollans@liu.se (L. Jollans), tino.ebbers@liu.se (T. Ebbers)

**Abstract**
Accurate segmentation of the heart is essential for personalized blood flow simulations and surgical intervention planning. Segmentations need to be accurate in every spatial dimension, which is not ensured by segmenting data slice by slice. Two cardiac computed tomography (CT) datasets consisting of 760 volumes across the whole cardiac cycle from 39 patients, and of 60 volumes from 60 patients respectively were used to train networks to simultaneously segment multiple regions representing the whole heart in 3D. The segmented regions included the left and right atrium and ventricle, left ventricular myocardium, ascending aorta, pulmonary arteries, pulmonary veins, and left atrial appendage. The widely used 3D U-Net and the UNETR architecture were compared to our proposed method optimized for large volumetric inputs. The proposed network architecture, termed Transformer Residual U-Net (TRUNet), maintains the cascade downsampling encoder, cascade upsampling decoder and skip connections from U-Net, while incorporating a Vision Transformer (ViT) block in the encoder alongside a modified ResNet50 block. TRUNet reached higher segmentation performance for all structures within approximately half the training time needed for 3D U-Net and UNETR. The proposed method achieved more precise vessel boundary segmentations and better captured the heart's overall anatomical structure compared to the other methods. The fast training time and accurate delineation of adjacent structures makes TRUNet a promising candidate for medical image segmentation tasks. The code for TRUNet is available at github.com/ljollans/TRUNet.

**INTRODUCTION**
Cardiac image segmentation is an important first step in the assessment of cardiac imaging data. Precise knowledge of cardiac morphology including the left ventricle (LV), left atrium (LA), left atrial appendage (LAA), aorta, and pulmonary veins (PV), allows for the extraction of quantitative clinical parameters, such as LV and LA volume as well as LV ejection fraction. Furthermore, accurate segmentations are also crucial for personalized cardiovascular fluid dynamics modeling, which shows great promise as a tool for assessing the risk of cardiovascular disease in patients with conditions such as atrial fibrillation [1]–[7]. The gold standard for defining cardiac geometries remains time-intensive manual segmentation of Computed Tomography (CT) or Magnetic Resonance Imaging (MRI) images by a skilled clinician. However, a growing number of automatic or semi-automatic deep learning methods for cardiac image segmentation are available. A 2017 whole-heart segmentation challenge found that segmentations were generally more precise when using CT compared to MRI data [8]. Due to high spatial resolution and contrast enhancement [9], [10], CT is optimally suited to precise anatomical delineation of cardiac structures.

The most commonly used deep learning method for medical image segmentation is the U-Net [11]–[15]. U-Net is a convolutional neural network (CNN) consisting of sequential encoding and decoding convolutions [16], [17]. The strength of the U-Net lies in its ability to derive low-level semantic features and combine them with high-level localization information using skip connections. Various augmentations for the U-Net have been proposed, including the addition of residual connections to speed up convergence and reduce overfitting [18], [19]. Several studies have used variants of the U-Net to automatically segment CT volumes of the heart. Sharkey and colleagues [20] used a self-configuring 3D U-Net trained on 100 CT volumes for whole-heart segmentation. The study found differences between the automatic segmentation and manual segmentations to be comparable to those seen between different clinical annotators, making their method a feasible alternative to manual segmentation. Using a small sample of 12 patients Bruns and colleagues [21] also showed that automatic deep learning segmentations can be reliably achieved across the whole cardiac cycle.

The U-Net and its variations, constrained by kernel size, typically only evaluate small patches of voxels at once, limiting the field of view of the network. The Vision Transformer (ViT) [22] extends the field of view by incorporating self-attention. Larger image patches are mapped to a lower-dimensional representation, thereby enabling long-range spatial dependencies. Incorporating ViT into the U-Net architecture, Chen et al. [23] developed *TransUNet* for 2D data, which applies the ViT module to the output of the final encoding block in a U-shaped network. Several transformer-based networks for 3D image volumes have also been proposed. These include several manuscripts (*UNETR* [24], *TransBTS* [25]) and peer-reviewed publications (*CoTr* [26], *MISSU* [27], *3D Brainformer* [28], *DR-LCT-UNet* [29]). Most of these methods maintain a U-shaped network structure but vary in how the ViT is incorporated. *UNETR* replaces the encoder cascade with a transformer block, the output of which is passed to the decoder cascade via the bottleneck and skip connections. In contrast, *TransUNet*, *CoTr*, *TransBTS* and *MISSU* add one or more Transformer blocks at the bottleneck, after a convolutional encoder cascade and before the decoder cascade. *3D Brainformer* and *DR-LCT-UNet* apply multiple transformer blocks, at each level of the encoder cascade and to the skip connections respectively, maintaining the U-shaped network structure.

These and other segmentation network architectures are customarily evaluated using instance segmentation tasks, such as labelling of tumours. Medical instance segmentation requires networks to recognize not only one or more lesions of different sizes and placements, but also in different tissues. *UNETR* [24] has been reported to perform less well on a brain tumour segmentation task [27], [28], [30]–[32] and on a coronary artery segmentation task [29] than other approaches including 3D U-Net, *TransUNet, MISSU* and *3D Brainformer*, although reported dice scores vary widely. *TransUNet* [23] and *Dual-swin TransUNet* [33] performed similarly to other methods for polyp [33] and brain tumour segmentation [27], although lower dice scores have also been reported [24], [34]. Compared with instance segmentation, simultaneous multi-label segmentation brings a different set of challenges [35]. Regions are typically adjoining, undivided, and in the same general arrangement in relation to each other. *TransBTS* reportedly achieved slightly higher dice scores than *UNETR* for a multi-organ segmentation task [36], while *UNETR* performed only slightly better than *TransUNet* and *CoTr* [24], [26].

The implementation of *TransUNet* proposed by Chen and colleagues [23] operates entirely in 2D but nevertheless achieves segmentation performance comparable to that of 3D approaches in instance segmentation and multi-label segmentation tasks [24], [27], [37]. A translation of *TransUNet* from 2D to 3D may further improve its efficacy. Parallel to the work reported here, another adaptation of *TransUNet* for 3D medical images was carried out [38]. The authors applied the network to the task of segmenting the LV cavity and myocardium from MRI and found dice scores to be higher for their approach than for other methods. At commencement of this project the code for that version of the network (Li et al., 2023) was not publicly available and was not examined in this study.

The following highlights the main contributions of this paper:
We implement and make available to other researchers a network architecture based on *TransUNet* and optimized for large volumetric imaging data.
By directly comparing the proposed network to the widely used 3D residual U-Net and *UNETR* we demonstrate the improved training speed and segmentation performance of our network.
Using publicly available data alongside a large sample of annotated cardiac CT volumes across the whole cardiac cycle we show that precise, fully automatic, simultaneous multi-label segmentation of the heart is possible with the proposed network architecture.

The remainder of this paper is organized as follows: Section II describes the network architecture. Section III describes the data used in this study. Section IV describes details of the implementation.

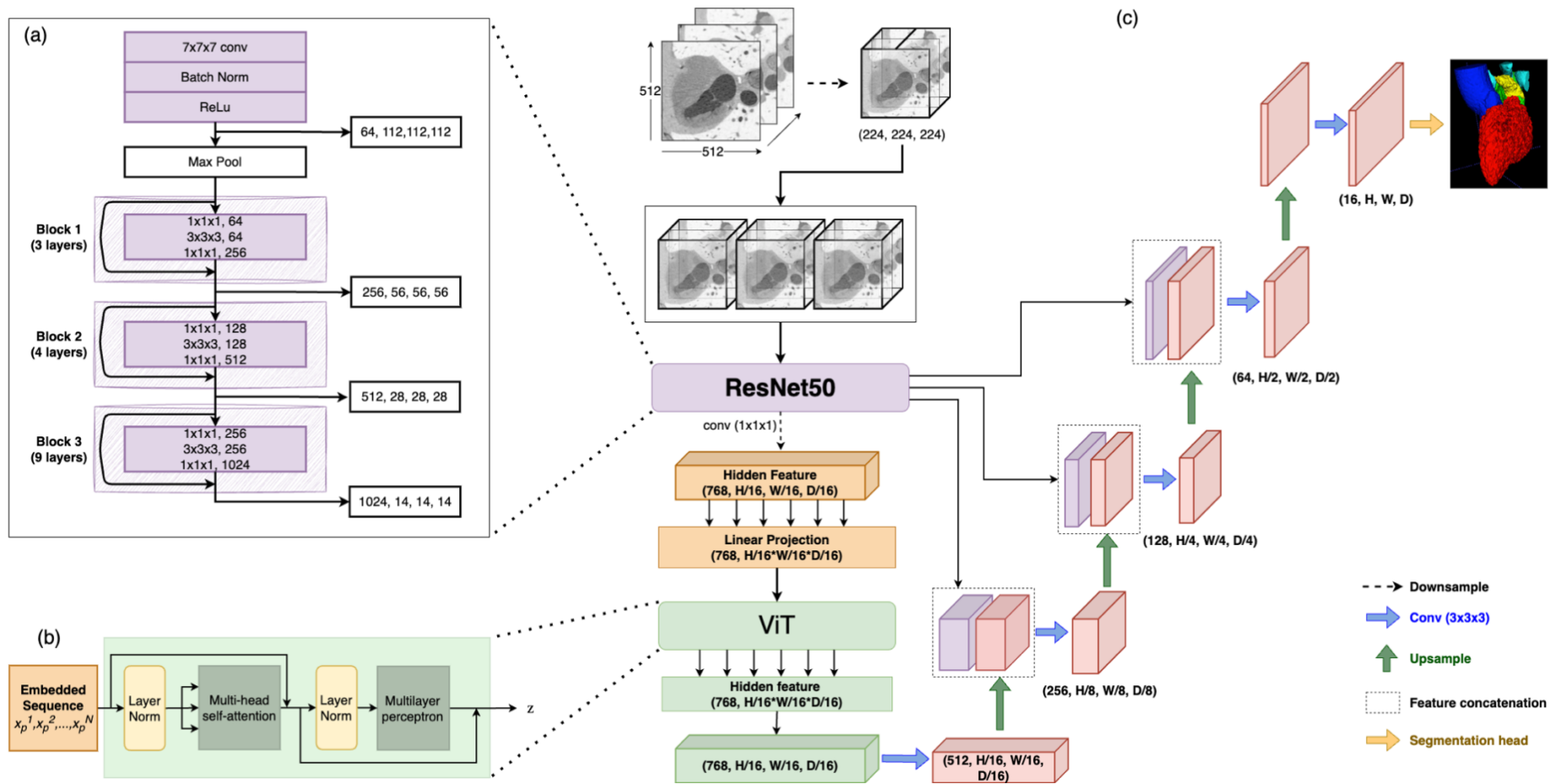

*Figure 1. Network architecture. (a) schematic of the modified ResNet50 block, (b) schematic of the Vision Transformer (ViT) block, (c) TRUNet encoder and decoder cascade with skip connections between the ResNet50 block and the decoder cascade.*

## I. Network architecture

The original *TransUNet* is composed of an encoder with three convolutional layers (64, 128 and 256 channels), a ViT block with 12 transformer layers, and a cascade upsampling decoder with four layers (256, 128, 64 and 16 channels) adding skip connections from the CNN encoder layers. A direct translation of this architecture to 3D resulted in a model with 918 Mio. parameters. Given hardware constraints it was not possible to train a model of this size. Chen et al. [23] also reported a variation of their network in which a modified ResNet-50 block was used as the encoder in place of the CNN block. While this variant did not perform as well as the original CNN-ViT encoder in their comparison it nevertheless showed promising results. Implementing the suggested modified ResNet-50 block resulted in a model with 134 Mio. Parameters.

Since the architecture of the encoder of the model tested here deviates from the CNN-ViT encoder used in *TransUNet* we will refer to this network architecture as Transformer Residual U-Net (*TRUNet*). The *TRUNet* architecture is shown in Figure 1. It is composed of an Encoder which includes the modified ResNet50 block, the ViT block, a cascade upsampling decoder, and a segmentation head.

**Encoder.** The input for the hidden layer of the ViT was generated using the hybrid encoder design described in [23]. In the first step the input data were embedded using a ResNet50 v2 variant. Since the output of the ResNet50 block was passed to the ViT with hidden layer size 768, the ResNet50 block was modified to generate output with size 1024 instead of 2048 by using 3 blocks consisting of 3, 4, and 9 layers instead of 4 blocks consisting of 3, 4, 6, and 3 layers. The output of the stem and the first two blocks of size 64, 256, and 512 respectively were passed to the decoder cascade via skip connections.

**Vision Transformer.** As in previous work [22], [23], [27] the patches used in the ViT were 16 pixels large along every axis. The input data size was therefore set to be 224*224*224 voxels. This resulted in a grid of 2744 non-overlapping patches representing 4096 voxels each. In line with ViT-Base described in [22], 12-layer multi-head attention with 12 heads was used. The size of the hidden layer was set to 768 and Multi-Layer Perceptron size to 3072. In place of three-dimensional positional embeddings, an array of zeros was added to the patch embedding since previous work has shown that ViT is capable of learning these [22], [23].

**Decoder.** A cascade upsampling decoder with 512, 256, 128 and 64 channels was employed. At each layer features were upsampled using trilinear interpolation with scaling factor 2, concatenated with the skip connection features, and passed through a 3*3*3 convolution and rectified linear unit activation function.

**Segmentation head.** The features were again trilinearly upsampled to the full resolution and a 3*3*3 convolution was used to reshape the output to the number of classes. Finally, a softmax activation function was used to generate the label probabilities.

Our proposed method was compared to two other network architectures implemented in MONAI 0.8.1 [39]. First, Residual U-Net (referred to from now on as *UNet*) based on [18] with 5 layers and 2-convolution residual units, convolution kernel size 3, stride 2, and Parametric Rectified Linear Unit activation. Second, *UNETR* as described in [24] with a Leaky Rectified Linear Unit activation function. For completeness we also trained *TRUNet* in

2D using slices along the z-axis. This corresponds to the R50-ViT Encoder and CUP Decoder framework outlined in [23].

## II. Data

**Internal CT dataset**
**Acquisition.** A total of 760 image volumes were used for training, validation, and testing. The image volumes represented time-resolved CT datasets from 39 patients which were part of a larger sample of cardiac CT data from routine medical ECG-gated coronary CT angiography examinations approved for retrospective research use. All datasets were acquired on a dual source 192*192 slice CT scanner (Siemens Somatom Force, Siemens Forchheim, Germany). Twenty time instances representing the whole cardiac cycle were acquired for each patient. Data were acquired at 512*512 voxel resolution with between 373 and 561 slices with 0.5mm thickness and 0.25mm overlap. Pixel size varied between 0.27 and 0.41mm (mean: 0.33, standard deviation: 0.034). For one patient all data were excluded and for a further two 13 and 7 volumes respectively were excluded due to poor image quality.

**Ground truth generation.** The ground truth was generated using semi-automatic and atlas-based segmentation. Multi-Atlas-based segmentations were obtained using the approach described in [40]. Hereby atlas is used to refer to a previously labelled image. Using registration, a transformation between a different target image and the atlas is estimated and the labels of the atlas are propagated into the target's image space. For multi-atlas segmentation the transformed labels from multiple atlases are combined into one segmentation.
The LA, LV, AA, PV, and LAA were segmented. Only the end-systolic and the end-diastolic timeframe were manually corrected and used to create atlases to automatically segment the remaining time instances. These time instances were selected to ensure that the most extreme changes in cardiac volume were correctly depicted. For eight patients the end-systolic timeframe was segmented semi-automatically in ITK-Snap [41] using the "Active contour" method to outline the blood pool and subsequent manual labelling and correction. These segmentations were used as atlases to segment the remaining volumes in the end-systolic phase with subsequent manual correction. For each patient the end-systolic segmentation was used as an atlas to segment the end-diastolic timeframe with subsequent manual correction. All segmentations for end-systole and end-diastole were visually inspected and approved by multiple readers.
Finally, the remaining timeframes were segmented automatically for each patient using the segmentations from end-systole and end-diastole as atlases. Visual inspection spot-checks of the automatic segmentations were carried out for each patient. This resulted in segmentations for two patients being manually corrected for two time instances each.

**MM-WHS Challenge Dataset**
CT data from the Multi-Modality Whole Heart Segmentation (MM-WHS) challenge [8] were used to externally validate the cardiac segmentation network. The annotation of these data included the LA, LV, AA, right atrium (RA), right ventricle (RV), left ventricular myocardium (Myo), and pulmonary artery (PA).

provided by the challenge organizers. Segmentations were evaluated qualitatively for the internal dataset by visual inspection of time-instances 1 and 11 for all patients in the test set.

## III.     Implementation

**Preprocessing**
To evaluate the effect of prior cropping of volumes to the cardiac ROI a localization network was trained using a binarized version of the multiclass labels. To train this localization network a *UNet* was trained using a subset of 14 patients from our internal dataset for 100 epochs with batch size 1. Dice score and visual inspection indicated an excellent identification of the ROI. For each patient, the bounding box was defined as the maximum of the coordinates produced by the localization network for each volume across the cardiac cycle. The bounding box was expanded by 20 voxels in each direction (corresponding to 10mm in z and between 5.39 and 8.16mm in x and y) to allow for error in the localization prior to cropping.
Volumes were downsampled to be 224 pixels large along every axis, in line with previous work using ViT [22], [23].
In each epoch the data were randomly rotated by up to 20° along a randomly selected axis with a likelihood of 50%, and randomly flipped along a randomly selected axis with a likelihood of 50%. Data loading times were optimized by applying cropping, transformations, and downsampling on the fly at training time.

**Network Training**
Analyses were conducted using Pytorch 1.9 [42] on a NVIDIA DGX A100 80GB GPU. The GPU memory limited the possible batch size to 2. The batch size for *TransUNet,* which was trained using slices, was set to 24.
Data were split into training, validation, and test sets at the patient-level prior to any analyses being carried out to keep these splits consistent across different methods. For the internal dataset this split was done semi-randomly to assign patients with fewer than 20 timepoints to the test set. 11 patients were assigned to the test set and 6 to the validation set. For the MM-WHS challenge dataset 40 volumes were pre-defined as the test-set. Of the 20 volumes in the training set 4 were used as the validation set.
The learning rate was initialized at 0.01, decaying by each iteration with power 0.9 using the poly learning rate strategy. The loss function used was the mean of Dice Loss and Cross Entropy Loss. The background was not included in the calculation of the loss function. Adam Optimizer was used for training [43].
Each network was trained for 72 hours, regardless of the number of iterations and epochs completed.
Post-processing involved removing small clusters: Only the largest connected component was retained in the segmentation for each ROI except the PV. For the PV all clusters that were at least half as large (in terms of voxel extent) as the largest PV component were retained.
For the internal dataset six networks were trained: *TRUNet*, *UNet*, and *UNETR* each with uncropped input volumes and with cropped input volumes. Based on results from the internal dataset *TRUNet*, *UNet*, and *UNETR* were trained only with uncropped input volumes for the MM-WHS challenge dataset.

**Evaluation**
For all analyses the performance the trained model which resulted in the highest dice score in the validation set was applied to the test set. Segmentation performance was evaluated quantitatively for the internal dataset using the dice score and $95^{th}$ percentile Hausdorff distance (HD95). The HD95 is sensitive to outliers and is therefore useful to assess global dissimilarity while the dice score captures local dissimilarities well.
For the challenge dataset the evaluation metrics (Dice score and Jaccard Index) were predefined in the evaluation script provided by the challenge organizers.
Segmentations were evaluated qualitatively for the internal dataset by visual inspection of time-instances 1 and 11 for all patients in the test set.

## IV. Results

**Internal dataset**

Within 72 hours runtime *TRUNet* completed 190 epochs (205 for *TRUNet*-c), *UNet* completed 251 epochs (303 for *UNET*-c), *UNETR* completed 198 epochs (219 for *UNETR*-c), and *TRUNet* 2D completed 74 epochs. Learning curves are shown in Figure 2.

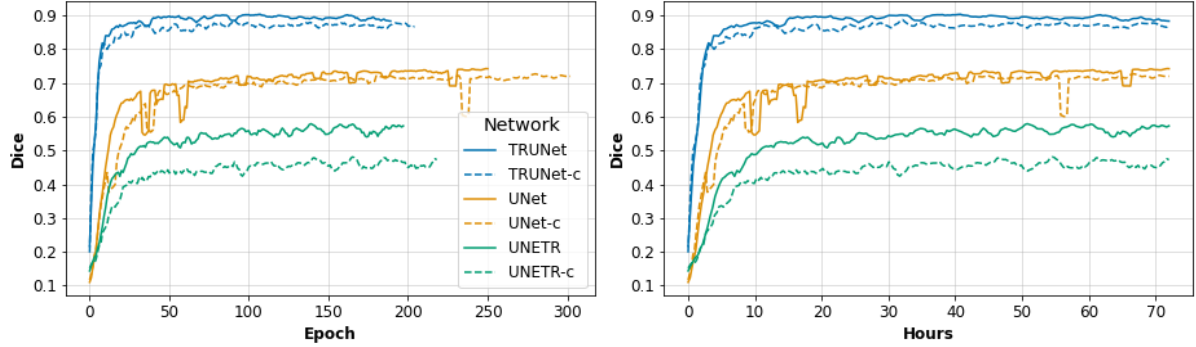

*Figure 2 Validation loss during training by Epoch and number of hours training for all networks. The best performing model was saved after around 36 hours (97 epochs) for TRUNet, 31 hours (89 epochs) for TRUNet-c, 71 hours (250 epochs) for UNet, 66 hours (279 epochs) for UNet-c, 72 hours (195 epochs) for UNETR, 54 hours (165 epochs) for UNETR-c, and 22 hours (24 epochs) for TRUNet 2D. TRUNet 2D is not shown since the validation set dice score during training is based on segmentation of slices not volumes, and values are therefore not comparable to the other methods.*

Overall performance was best for *TRUNet* (see dice scores in Table 1 and HD95 in Table 2). Dice scores for the LV and LA were highest for *TRUNet*-c and dice score for PV was highest for *UNet*. HD95 for LV was lowest for *TRUNet*-c.

Two four-way ANOVAs found that network architecture was the main source of variation in dice score ($F_{Dice}(2, 6000) = 1794.32$, $p < 0.0000001$) and in HD95 ($F_{HD95}(2, 6000) = 921.82$, $p < 0.0000001$). The main effect of whether cropping was used ($F_{Dice}(1, 6000) = 53.65$, $p = 2.71e-13$; $F_{HD95}(1, 6000) = 64.74$, $p = 1.02e-15$), patient ($F_{Dice}(10, 6000) = 58.54$, $p = 2.69e-37$; $F_{HD95}(10, 6000) = 16.60$, $p = 9.67e-11$), and time instance ($F_{Dice}(19, 6000) = 4.97$, $p = 0.0019$; $F_{HD95}(19, 6000) = 13.69$, $p = 6.71e-09$) were also significant. For both dice score and HD95 there were significant interaction effects between model architecture and use of cropping ($F_{Dice}(2, 6000) = 11.17$, $p = 1.44e-05$; $F_{HD95}(2, 6000) = 12.13$, $p = 5.54e-06$), between patient and model architecture ($F_{Dice}(20, 6000) = 5.69$, $p = 6.90e-08$; $F_{HD95}(20, 6000) = 7.75$, $p = 2.13e-10$) and between patient, use of cropping, and network architecture ($F_{Dice}(20, 6000) = 12.03$, $p = 7.52e-21$; $F_{HD95}(20, 6000) = 5.36$, $p = 6.32e-08$). For dice score there was also a significant interaction between patient and use of cropping ($F_{Dice}(10, 6000) = 7.37$, $p = 6.36e-04$). For HD95 there were significant interaction effects between model architecture, patient, and time instance ($F_{HD95}(380, 6000) = 1.49$, $p = 9.04e-08$), between model architecture and time instance ($F_{HD95}(38, 6000) = 2.15$, $p = 5.88e-04$) and between patient and time instance ($F_{HD95}(190, 6000) = 2.41$, $p = 2.90e-11$). These effects were found to be driven by results for one volume for which *UNETR*-c failed to segment any of the ROIs.

Visual inspection of segmentations generated by *UNETR* revealed that the segmentations were highly fragmented. While the blood pool was segmented adequately in all but one inspected volume the labelling of the ROIs was poor. Visual inspection of the

*UNet* and *TRUNet* segmentations revealed that lower dice scores for the LAA and PV compared to other ROIs were due to variation in the placement of the boundary between the LA and these ROIs, as well as the point at which the segmentation in the PV terminates. The exact placements of these boundaries, while varying from the ground truth, were determined to fall within an acceptable margin of error in most inspected segmentations.

Either the boundary between the LV and AA or between the LV and LA were poorly segmented in four out of 11 patients by *UNet* (four partially overlapping patients for *UNet*-c). LV and AA/LA boundary segmentations were poor in two patients for *TRUNet*-c and in none for *TRUNet*. ROIs were undersegmented in all patients by *UNet* (18 volumes) and *UNet*-c (19 volumes). *TRUNet*-c undersegmented ROIs in 11 volumes from eight patients and *TRUNet* undersegmented ROIs in six volumes from five patients. *TRUNet* mainly undersegmented the LAA (in five volumes from four patients). For one of these patients the LAA was not fully included in the image volume, making it impossible to place the LAA orifice accurately in the ground truth segmentation.

*Table 1. Test Set dice score*

|  | Left Ventricle | Left Atrium | Left Atrial Appendage | Ascending Aorta | Pulmonary Veins | Average* |
|---|---|---|---|---|---|---|
| *UNet* | 0.9495 | 0.9507 | 0.8393 | 0.9559 | **0.7329** | 0.8857 |
| *UNet-c* | 0.9480 | 0.9424 | 0.7849 | 0.8382 | 0.7292 | 0.8485 |
| *TRUNet* | 0.9529 | 0.9531 | **0.8614** | **0.9687** | 0.7196 | **0.8911** |
| *TRUNet-c* | **0.9545** | **0.9542** | 0.8531 | 0.9448 | 0.7248 | 0.8863 |
| *UNETR* | 0.8221 | 0.6757 | 0.7160 | 0.7004 | 0.6127 | 0.7054 |
| *UNETR-c* | 0.7688 | 0.6091 | 0.6757 | 0.6134 | 0.5839 | 0.6502 |
| *TRUNet 2D* | 0.9430 | 0.9413 | 0.8344 | 0.9668 | 0.6876 | 0.8746 |

*Macro-Average calculated without the background*

*Table 2. Test Set HD95*

|  | Left Ventricle | Left Atrium | Left Atrial Appendage | Ascending Aorta | Pulmonary Veins | Average |
|---|---|---|---|---|---|---|
| *UNet* | 8.68 | 12.32 | 9.09 | 8.33 | 18.61 | 11.40 |
| *UNet-c* | 10.36 | 19.46 | 20.32 | 37.40 | 19.91 | 21.49 |
| *TRUNet* | 7.14 | **9.02** | **7.71** | **4.29** | **15.10** | **8.65** |
| *TRUNet-c* | **6.75** | 9.84 | 8.52 | 6.95 | 17.21 | 9.85 |
| *UNETR* | 73.66 | 52.75 | 18.10 | 82.03 | 23.38 | 49.99 |
| *UNETR-c* | 67.12 | 72.54 | 21.60 | 120.59 | 28.21 | 62.01 |
| *TRUNet 2D* | 15.18 | 14.90 | 11.97 | 6.97 | 17.26 | 13.25 |

*Macro-Average calculated without the background*

**MM-WHS CT data**

The highest validation set dice score was reached by *TRUNet* after 507 of 1122 total completed epochs (approximately 30 hours), by *UNet* after 907 of 1229 total completed epochs (after approximately 52 hours), and by *UNETR* after 1191 of 1192 total completed epochs (approximately 72 hours). Results are shown in Table 3 and 4. Dice score and Jaccard index were highest for *TRUNet* for all ROIs.

Table 3. Dice score for MM-WHS challenge

|                  | *TRUNet* | *UNet* | *UNETR* |
|------------------|----------|--------|---------|
| Myocardium       | **0.8710** | 0.8372 | 0.7766 |
| Left Atrium      | **0.8385** | 0.7961 | 0.8131 |
| Left Ventricle   | **0.8458** | 0.7013 | 0.7216 |
| Right Atrium     | **0.9184** | 0.8116 | 0.7260 |
| Right Ventricle  | **0.8612** | 0.7414 | 0.6427 |
| Ascending Aorta  | **0.9172** | 0.7821 | 0.6951 |
| Pulmonary Artery | **0.7947** | 0.6560 | 0.6070 |
| Average*         | **0.8753** | 0.7811 | 0.7438 |

*Macro-average calculated including the background

Table 4. Dice score for MM-WHS challenge

|                  | *TRUNet* | *UNet* | *UNETR* |
|------------------|----------|--------|---------|
| Myocardium       | **0.7925** | 0.7352 | 0.6612 |
| Left Atrium      | **0.7285** | 0.6714 | 0.6912 |
| Left Ventricle   | **0.7542** | 0.5735 | 0.5803 |
| Right Atrium     | **0.8518** | 0.7107 | 0.6177 |
| Right Ventricle  | **0.7630** | 0.6123 | 0.5024 |
| Ascending Aorta  | **0.8523** | 0.6876 | 0.5917 |
| Pulmonary Artery | **0.6829** | 0.5294 | 0.4780 |
| Average*         | **0.7819** | 0.6529 | 0.6061 |

*Macro-average calculated including the background

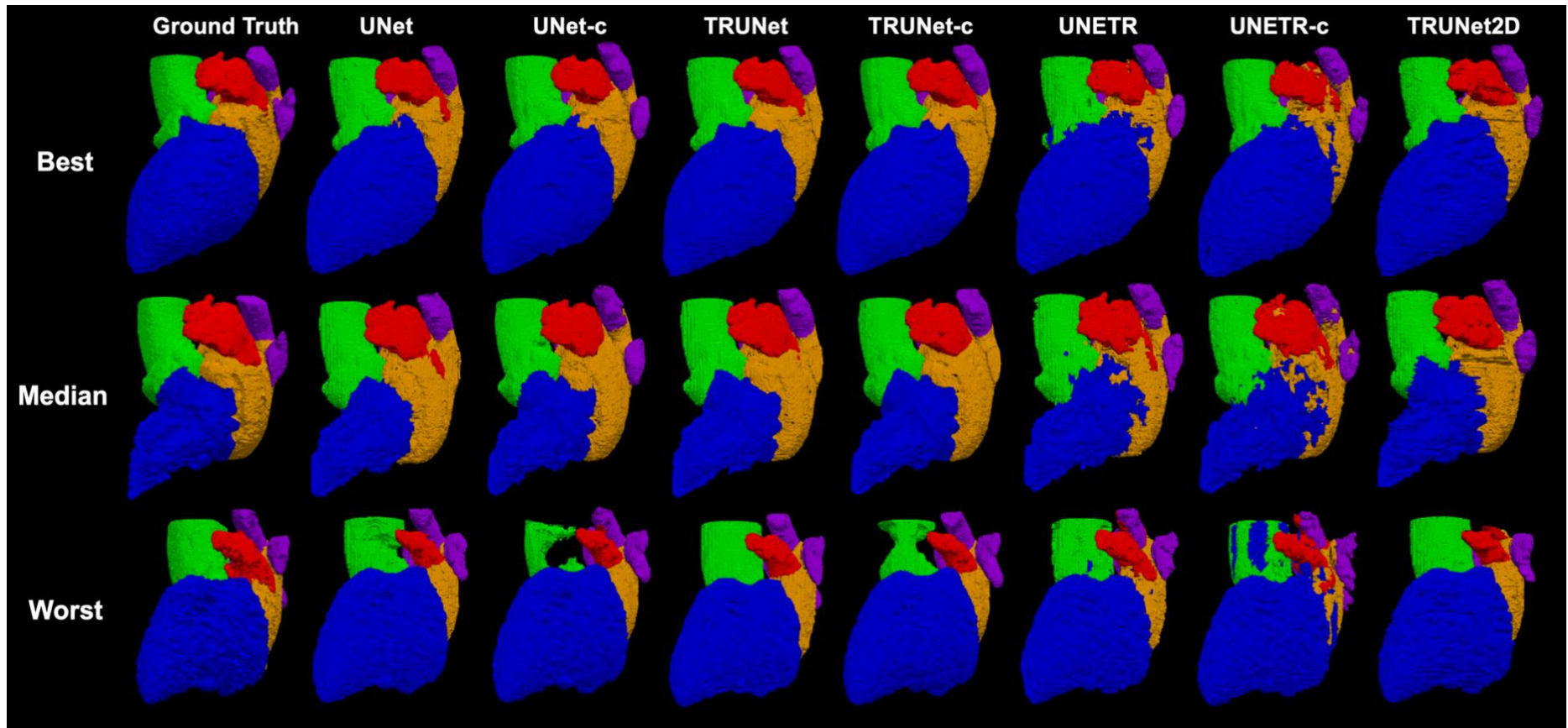

*Figure 3 Best, median, and worst segmentations as defined by joint ranking of average dice score and HD95 across all methods. The best and median segmentations are of the same patient at mid systole and end systole respectively. The worst segmentation is of another patient at end diastole.*

## V. Discussion

In the present study we implemented a novel Vision Transformer-based method for segmentation of 3D cardiac CT image volumes. Our method, termed *Transformer Residual U-Net* (*TRUNet*), is based on *TransUNet* [23], modified for 3D inputs and adapted to accommodate large-scale imaging data under computational constraints. Using two separate datasets we showed that *TRUNet* converged in close to half the time needed by other methods while also outperforming Residual 3D U-Net [18] and *UNETR* [24] in terms of segmentation performance.

The ViT provides the advantage of capturing the global context differently than convolutional layers by flattening the image to a linear projection. Many studies have highlighted that networks incorporating ViT modules are very well suited to segmentation of medical images and image volumes [24], [27], [28], [30]–[34], [36]–[38]. The primary difference between *UNETR* and Trans*UNet* is the encoder. While Trans*UNet* uses a hybrid cascade downsampling CNN and ViT encoder, *UNETR* does not include a cascade downsampling encoder. In this study we found that a 3D *UNet* without a ViT block was able to produce better segmentations in a simultaneous multi-label segmentation task than *UNETR*. This indicates that the cascade downsampling encoder design is crucial to producing good multi-label segmentations. The cascade downsampling approach alongside skip connections enables the network to leverage information at multiple scales - from local to global - at once. Our proposed framework, which uses a hybrid ResNet50-ViT encoder, outperformed both 3D *UNet* and *UNETR* in the multi-label segmentation task. The addition of the ViT block can be assumed to provide additional global image context. Our finding that incomplete segmentation of regions was less prevalent for *TRUNet* than for the other frameworks supports this conclusion. Our findings not only demonstrate that leveraging the context of the whole image volume through the ViT block is beneficial, but also that reducing image context by cropping is likely counterproductive. Cropping of image volumes to the cardiac region increased the rate of misclassifications, as seen by visual inspection and evaluation of HD95 values.

The main source of error which we observed for *TRUNet* was undersegmentation of the LAA. Accurate segmentation of the LAA from CT is a challenging task since the boundaries of the structure are often not obvious to the human eye. The shape of the LAA varies widely between patients [44] and is known to play an important role in thrombus formation [45]. However, the LAA is typically not included as a region of interest in segmentations of the heart. The LAA is either explicitly excluded [8], [20], included as part of the LA region [14], or how the LAA is treated is not specified [21]. Previous attempts to automate LAA segmentation in CT required manual input to define a bounding box [46], [47]. In MRI, a fully automated atlas-based approach reached a dice score of 0.91 for the LAA [48]. To the best of our knowledge similarly good automatic segmentations of the LAA have not been achieved in CT. However, precise LAA segmentation is critically important in clinical cardiology for tasks such as planning surgical procedures [47]. *TRUNet* accurately placed the LAA in all inspected volumes and segmented the region more accurately than the other methods. However, *TRUNet* was not able to capture all details of the structure's shape.

The highly individual variations in LAA shape and its small size may require a different training approach and would benefit from increased training set size.

*TRUNet* trained using image slices instead of volumes was able to produce segmentations of relatively high quality compared to the 3D approaches. While segmentations created in 2D are likely to have some degree of slice artifact when concatenated into three dimensions, the use of slices rather than volumes for training is a feasible alternative for circumstances where memory constraints make training in 3D difficult. The markedly faster time to convergence for *TRUNet* also makes this network architecture a promising candidate for use in resource-constrained environments. While some recent work has highlighted differences in computational overhead and runtime between deep learning networks for medical image segmentation [27], factors such as training time and memory requirements of complex network architectures are often not reported. For *UNETR* we found that segmentation performance continued to increase throughout training. *UNETR* may therefore have improved to a similar level as *TRUNet* with increased training time. The same is likely also true for 3D *UNet*. This is supported by higher dice and jaccard scores of similar networks trained using the same data in the MM-WHS challenge [8]. While *TRUNet* achieved metrics in line with the average for other deep learning methods, 3D *UNet* performed markedly worse in the present study than in previous work. Where training times were reported, networks were trained for 30000 iterations compared to around 16000 iterations for our model [49], [50]. Higher segmentation performance of previously reported approaches may also be due to additional steps taken to improve results. These included enlargement of the training set [51] and consideration of shape context [52].

That our segmentation attempt of a publicly available dataset with a similar framework yielded worse results than previously reported work using 3D *UNet* is one of the main limitations of this work. It is possible that continued training and different data processing steps would have increased performance of the *UNETR* and 3D *UNet* models. The design of this study, which was intended to directly compare performance of the network architectures given the same ressources and conditions, limits the extent to which our findings using the MM-WHS dataset can be directly compared to other work using this data. Nevertheless, within the scope of this work the *TRUNet* architecture was found to produce significantly superior multi-label segmentations than either 3D *UNet* or *UNETR*. Furthermore, the CNN encoder recommended by the Trans*UNet* authors. [23] could not be implemented here due to memory limitations while working with 3D data. Given sufficient computational resources, a direct comparison of the ResNet50 block used in *TRUNet* and the CNN block used in Trans*UNet* would be beneficial. A possible modification for *TRUNet* is the incorporation of Swin transformers which would reduce computational complexity while producing similar results [33]. Finally, the input volume size constraint given by the ViT block limits the type of data this network can be used with. In future work we intend to adapt *TRUNet* to accept other input dimensionalities while minimizing information loss caused by downsampling.

Future work should focus on testing the *TRUNet* cardiac segmentation model in common conditions such as congenital heart disease [53], and on evaluating the capacity of the model to be retrained with new patient data. Our proposed network architecture was able to produce high quality multi-label segmentations of the heart from CT volumes within less than 15 hours of training time, making it a good candidate for training with limited time or computational ressources. Conversely, while the size of the trained model is much larger for *TRUNet* than for the other models, inference for new volumes of took less than 30 seconds for volumes with 512*512 matrix resolution and less than 60 seconds for volumes with 1024*1024 resolution. While slower, inference can also be run using CPU and was tested with several different GPUs.

## VI. Conclusion

In this paper we have shown that a cascade downsampling encoding scheme is of great value for simultaneous multi-label segmentation of medical image volumes in 3D. Inclusion of a ViT block in the encoder was found to not only greatly speed up training but also improve the quality of segmentations. Using two separate datasets we showed that our proposed network architecture (*TRUNet*) is able to accurately segment the left side of the heart across the whole cardiac cycle and produces good whole-heart segmentations with only 16 CT volumes for training. Although segmentations of the left atrial appendage using this model do not fully capture inter-individual variation in the anatomy of the appendage, *TRUNet* outperformed residual 3D UNet and UNETR in segmentation of all regions of interest and in the placement of the boundaries between the left ventricle and ascending aorta and left atrium.

Acknowledgements:

The computations and data handling were enabled by resources provided by the National Academic Infrastructure for Supercomputing in Sweden (NAISS) at Linköping University partially funded by the Swedish Research Council through grant agreement no. 2022-06725.

We wish to thank Andre Da Luz Moreira and Sophia Bäck for their role in acquiring the data used in this study and for managing the segmentation process. Many thanks to Norman Janurberg, Christian Luksitch, and Aman Kumar Nayak for their work generating ground truth segmentations with the guidance of Andre Da Luz Moreira. Faisal Zaman supplied the registration tool used in the ground truth segmentation, for which we are very grateful. We also wish to thank Sophia Bäck for lending her expertise in cardiac anatomy to assist in judging the quality of the automatically generated segmentations. Many thanks also to Jonas Lantz for his help in accessing computational infrastructure and assistance in visualization of our results. Lastly, we wish to thank Erik Ylipää and the AIDA team at Linköping University for valuable methodological discussions.

# Appendix

## One-dimensional positional embedding

When passed a one-dimensional positional embedding, *TRUNet* completed 208 epochs within 72h. The highest dice similarity score (dice score) was achieved at epoch 166 at 57.04 hours. dice score and 95$^{th}$ percentile Hausdorff distance (HD95) in the test set overall and for each ROI are reported in Table S1.

*Table S1. Dice Similarity Score (dice score) and 95$^{th}$ percentile Hausdorff Distance (HD95) overall and for each ROI for TRUNet when passed a one-dimensional positional index.*

|  | Left Ventricle | Left Atrium | Left Atrial Appendage | Ascending Aorta | Pulmonary Veins | Average |
|---|---|---|---|---|---|---|
| *dice score:* |  |  |  |  |  | 0.8649 |
| *Mean* | 0.9405 | 0.9207 | 0.8196 | 0.8748 | 0.7688 | (0.1047) |
| *SD* | (0.0366) | (0.0674) | (0.1210) | (0.1365) | (0.0626) |  |
| *HD95: Mean* | 36.10 | 47.52 | 32.87 | 72.55 | 15.22 | 40.86 |
| *SD* | (53.06) | (57.29) | (54.49) | (114.36) | (5.44) | (65.64) |

## Removal of small clusters

In order to explore the change in segmentation performance when only the largest components were retained for each region of interest, a post-processing step was added to remove small clusters of voxels. For the Left Ventricle (LV), Left Atrium (LA), Left Atrial Appendage (LAA) and Ascending Aorta (AA) only the largest connected component was retained. For the Pulmonary veins (PV) a variable number of components were retained since the ground truth segmentations also contained multiple components. All clusters that were at least half as large (in terms of voxel extent) as the largest PV component were retained. dice score values are shown in Table S2. HD95 values are shown in Table S3.

For *UNet*-c some of the missegmented clusters were larger than clusters in the intended ROI. This resulted in the LAA being completely missed in the segmentations for 8 volumes from 2 patients (out of 200 total volumes) and the AA being completely missed in 17 volumes from 2 patients when only the largest cluster was retained. Only retaining the largest cluster for each ROI did not lead to gross missegmentations by any of the other networks but did result in reduced dice score due to disconnected components being removed. In the case of one patient where the LAA was hard to segment and not fully included in the volume, removing smaller clusters resulted in lower dice score for all methods.

*Table S2. Average (SD) Dice Similarity Score overall and per ROI for all Networks when only large components are retained in the segmentation.*

|  | Left Ventricle | Left Atrium | Left Atrial Appendage | Ascending Aorta | Pulmonary Veins | Average |
|---|---|---|---|---|---|---|
| *UNet*-c | 0.9480 (0.0317) | 0.9424 (0.0318) | 0.7849 (0.1975) | 0.8382 (0.2792) | 0.7292 (0.0819) | 0.8485 (0.0795) |
| *UNet* | 0.9495 (0.0176) | 0.9507 (0.0168) | 0.8393 (0.0954) | 0.9559 (0.0233) | **0.7329** (0.0715) | 0.8857 (0.0255) |
| *TRUNet*2D | 0.9430 (0.0243) | 0.9413 (0.0202) | 0.8344 (0.0743) | 0.9668 (0.0153) | 0.6876 (0.0981) | 0.8746 (0.0276) |
| *TRUNet*-c | **0.9545** (0.0159) | **0.9542** (0.0195) | 0.8531 (0.0907) | 0.9448 (0.0756) | 0.7248 (0.0756) | 0.8863 (0.0291) |
| *TRUNet* | 0.9529 (0.0179) | 0.9531 (0.0159) | **0.8614** (0.0835) | **0.9687** (0.0144) | 0.7196 (0.0725) | **0.8911** (0.0215) |

*Table S3. Average (SD) 95$^{th}$ percentile Hausdorff Distance overall and per ROI for all Networks when only large components are retained in the segmentation.*

|  | Left Ventricle | Left Atrium | Left Atrial Appendage | Ascending Aorta | Pulmonary Veins | Average |
|---|---|---|---|---|---|---|
| *UNet*-c | 10.36 (13.60) | 19.46 (21.10) | 20.32 (45.92) | 37.40 (88.51) | 19.91 (10.76) | 21.49 (21.46) |
| *UNet* | 8.68 (6.14) | 12.32 (16.69) | 9.09 (5.80) | 8.33 (8.82) | 18.61 (7.41) | 11.40 (4.77) |
| *TRUNet*2D | 15.18 (14.13) | 14.90 (9.27) | 11.97 (8.18) | 6.97 (9.38) | 17.26 (6.79) | 13.25 (5.04) |
| *TRUNet*-c | **6.75** (4.33) | 9.84 (9.61) | 8.52 (4.82) | 6.95 (8.39) | 17.21 (7.62) | 9.85 (4.13) |
| *TRUNet* | 7.14 (5.15) | **9.02** (3.35) | **7.71** (3.24) | **4.29** (6.21) | 15.10 (6.43) | **8.65** (2.47) |